\documentclass[sigconf]{acmart}
\usepackage{todonotes}
\usepackage{amsmath}
\usepackage{float}
\AtBeginDocument{%
  \providecommand\BibTeX{{%
    \normalfont B\kern-0.5em{\scshape i\kern-0.25em b}\kern-0.8em\TeX}}}

\acmConference[CTHCI Project]{Current Topics in HCI - Project}{July 10, 2019}{Aachen, DE}
  
\begin{document}

\setlength{\belowcaptionskip}{-10pt}

\title{Action Bar Adaptations for One-Handed Use of Smartphones}
\author{Siddharth Mehrotra, Saurav Das, and Sourabh Zanwar}
\affiliation{%
  \institution{RWTH Aachen University}
  \streetaddress{Templergraben 55}
  \city{Aachen}
  \state{North Rhine-Westphalia}
  \country{Germany}
  \postcode{52062}
}
\email{firstname.lastname@rwth-aachen.de}

\begin{abstract}
One-handed use of smartphones is a common scenario in daily life. However, use of smartphones with thumb gives limited reachability to the complete screen. This problem is more severe when targets are located at corners of the device or far from the thumb's reachable area. Adjusting screen size mitigates this issue by making screen UI to be at the reach of the thumb. However, it does not utilize available screen space. We propose UI adaptation for action bar to address this. With our results, designed adaptations are faster for non-dominant hand and provides significantly better grip stability for holding smartphones. Intriguingly, users perceived our system as faster, more comfortable and providing safer grip when compared with the existing placement of action bar. We conclude our work with video analyses for grip patterns and recommendations for UI designers.
\end{abstract}

\begin{CCSXML}
<ccs2012>
<concept>
<concept_id>10003120.10003121.10003122.10003334</concept_id>
<concept_desc>Human-centered computing~User studies</concept_desc>
<concept_significance>500</concept_significance>
</concept>
<concept>
<concept_id>10003120.10003121.10003124.10010865</concept_id>
<concept_desc>Human-centered computing~Graphical user interfaces</concept_desc>
<concept_significance>500</concept_significance>
</concept>
<concept>
<concept_id>10003120.10003121.10003128</concept_id>
<concept_desc>Human-centered computing~Interaction techniques</concept_desc>
<concept_significance>500</concept_significance>
</concept>
</ccs2012>
\end{CCSXML}

\ccsdesc[500]{Human-centered computing~User studies}
\ccsdesc[500]{Human-centered computing~Graphical user interfaces}
\ccsdesc[500]{Human-centered computing~Interaction techniques}

\keywords{UI adaptation; one-handed; reachability; smartphone}


\maketitle

\section{Introduction}

\begin{figure}[h]
\centering
  \includegraphics[angle=90,width=1.0\columnwidth]{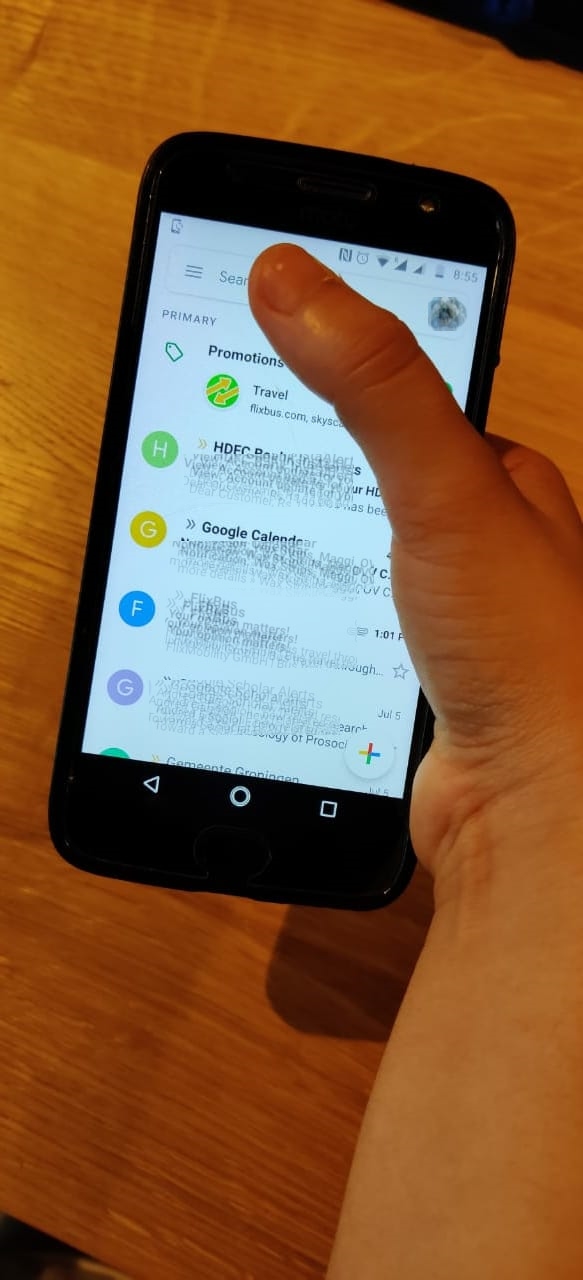}
   \caption{A user using the action bar on an Android phone with the GMail App.}~\label{fig:android-user}
\end{figure}

Smartphone screens in today's world have increased considerably in size. Looking back at the rise of the smartphone era with Apple iPhone coming to the market in 2007 with dimensions (115 x 61 x 11.6 mm) all the way to the latest iPhone, iPhone XR with dimension (150.9 x 75.7 x 8.3 mm), we can already see how the phones have become taller and thinner. Smartphones with bigger screens are now popular owing to the fact that they are becoming one of the primary multimedia content viewing devices in our everyday lives. With more screen area displaying more content, problems such as reachability are becoming predominant. The reachability issue occurs since most of the users use smartphones in one-handed mode \cite{accMobDevices} rather than in the cradled or two-handed modes in which reachability issues are far less. This choice of holding the phone may be attributed to the fact that usually users are multi-tasking and the other hand is being used to perform some low cognition task or tasks involving only procedural knowledge. In one-handed mode the screen is accessed by the thumb only and it's range and flexibility is limited. Usually users cope with this by tilting the device to bring the far screen corner closer to the stretched thumb \cite{chang2015understanding, negulescu2015grip} but this technique has limited scope and also compromises the grip stability.

Most common technique used by users to deal with reachability is to change the gripping pattern in order to reach far targets. Also, with recent advancements the phone layout can be changed to bring the top panel to the center of the phone. However, we believe by doing this the benefits of a large screen is not being fully utilized. 

We propose UI Adaptation for smartphones in one-handed mode, a method in which we address the issue of reachability of existing UI elements. We use the Orientation information from the device to make changes in the UI to better position elements across the screen. We use the UI element action bar because it is one of most common UI element across all apps, offers change in grip for one hand usage (because of it's position) and it is essential for visually challenged users as screen readers starts from action bars. We document time taken, orientation angles, and our observations related to various grip patterns \& gripping strategies that users take to cope with reachability. We conclude with recommendations for inclusion of our technique in Smartphone Applications.  
In summary, this paper makes the following contributions:
\begin{enumerate}
\item We compare 3 different adaptations of action bar based on existing UI element for touchscreen mobile phones.
\item a quantitative study examining the effect of different phone sizes and interaction hand on UI adaptations.
\item Empowering our results based on user's Likert scale responses and video analyses of the experiment to study the overall experience with the adaptations.
\end{enumerate}

\section{Related Works}
There are variety of techniques in the commercial market for tackling reachability in smartphones. With Apple's iPhone users can turn on \textit{Reachability} to bring items at the top of the screen down to the lower half of the screen\footnote{https://support.apple.com/guide/iphone/reachability-iph66e10a71c/ios}. With Motorola swipe to shrink screen mode, user can make screen smaller for one-handed use \footnote{https://motorola-global-portal.custhelp.com}. In addition, there are techniques to tweak the keyboard layout for one-handed operation. The keyboard snaps to one side of the screen and one can use the arrow button to switch sides\footnote{https://gizmodo.com/the-essential-guide-to-using-your-oversized-phone-with-1830709197}. However, re-sizing the screen outcasts the benefit of the large screen smartphones \cite{kim2012interaction}.

A good amount of literature already exists focusing on one-handed use of devices \cite{girouard2015one}\cite{huot2006spiralist}. Lehtovirta \& Oulasvirta has modelled the functional area of the thumb on mobile touchscreens. Their model ensures that a user interface is suitable for interaction with the thumb \cite{bergstrom2014modeling}. Chang et. al in CHI'15 constructed a design space for one-handed targeting interactions. They conducted an empirical experiment to discover usage patterns of tilting devices toward user thumb for touch screen regions \cite{chang2015understanding}. Hong and Lee introduced \textit{TouchShield} which provides shortcuts to frequently used commands via the thumb. The designed virtual control (\textit{TouchShield}) provides an area in which the thumb can pin the phone down in order to provide a stable grip \cite{hong2013touchshield}. Although these methods tackles reachability issues and grip stability, they introduce new concepts which needs to be adapted first as indirect inputs and requires additional use of thumb's press for hitting the target.

In another attempts to tackle reachability issue from existing UI perspective, Alt \& Buschek in CHI'17 proposed \textit{ProbUI}, a model that automatically evaluate user's touch sequences. They intend to guess the behaviour and target of the user based on machine learning probabilistic models \cite{buschek2017probui}. Karlson \& Bederson introduced \textit{ThumbSpace} and \textit{Shift} ---two software based interaction techniques. \textit{ThumbSpace} addresses distant objects while \textit{Shift} addresses small object occlusion. Their results supports better overall speed and accuracy in hitting small targets when combining both the techniques \cite{karlson2008one}.

While such methods can be used to predict the user's intention for button press however the reachability issue is still an open question as generative models have to surpass multiple UI elements before reaching targets on screen corners. Using a similar approach, Alt et. al introduces trigger based dynamic change in layout for rolling bar and floating menu button. Their results shows that trigger based elements improve reachability on a large
device, and can reduce interaction time and device tilting \cite{buschek2017dynamic}. Our research contribution is different from these prior works in the form that adapting existing UI element and measuring overall device orientation for phone stability. We focus on re-orienting the device only when it is required rather than based on triggers which is not ideal in real life context. In addition, we include comparison across hand dominance for wider audience reach.

\section{UI Adaptation Technique}

The UI Adaptation technique in Smartphones is a technique which can mitigate certain aspects of the reachability issue. Anchoring and Repositioning are known methods used to handle application screens at different device orientations. With UI Adaptation, the individual UI elements have a not-too-strict layout. The UI elements can be categorically segregated according to types, frequency or importance to the user at a particular context. For items lying in the screen space out of the thumb's reach, these adaptations will rearrange the UI based on the \textit{reach-intent} of the user. 

The term \textit{reach-intent} is defined as the intention of the user to click a specific area of the Application while tilting the device in single handed mode. Device Orientation coupled with categorization of UI elements on a loosely based layout gives us a way to guess the UI element user wants to access and make them more reachable.

\section{User Study}

In order to understand whether UI adaptation is a useful technique, we wanted to study the target selection speed and device orientation angles in the case where regular UI elements are re-positioned. We took a common UI element, the Action Bar and carried an User Study with 8 participants (5 males and 3 females, 21-29 years, M = 25.39, SD = 2.7). All our users were right-handed with about 8-9 years of experience with smartphone interaction. The Action Bar position was varied in 2 different ways in addition to the usual top position (the control condition here). The Action Bar area was divided into 6 sub-areas marked from 1-6. This was done to understand the difference in measurements for the far and the near target areas. The buttons were all equally sized for all the UI orientations. The sequence in which the buttons need to be pressed came as an instruction message in the screen along with information about which hand the user should use when clicking the button. 

\begin{figure}[h]
\centering
  \includegraphics[width=1.0\columnwidth]{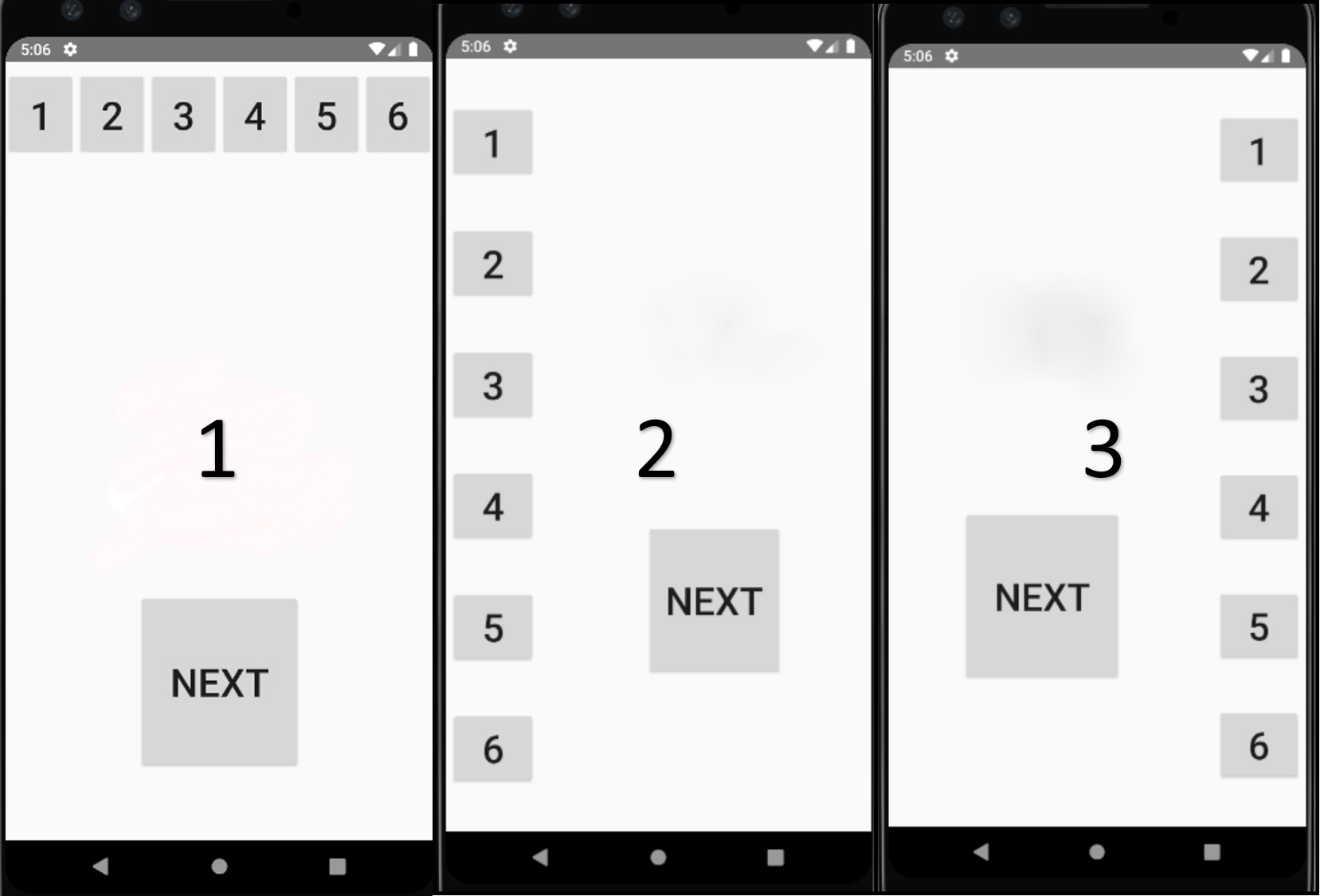}
   \caption{UI of our study prototype showing the baseline and adapted positioning of the action bar. Left: UI with number 1 is the baseline 'top' layout of the action bar. Right: UI with number 2 \& 3 are adaptation 'left' and 'right' layout of the action bar.}~\label{fig:our-app}
\end{figure}

The sequence was preset in the application by randomizing a number sequence before. The participants were asked to select numbered buttons on the screen according to the number shown on the instruction portion of the screen. This had to be done for all 3 Action Bar placements and while using the dominant and the non-dominant hand in turn. The participants were asked to always use the device in one-handed mode, in portrait orientation and without supporting their arm in any way while using the device.

We recorded the overall time taken by the users to hit the target once a number corresponding to 6 sub-areas displayed at the center of the screen. Once all 6 sub-areas were pressed the user was asked to click on 'next' button for another UI layout. The size of the buttons on sub-areas were kept similar to actual guidelines by Android Community \footnote{https://material.io/design/components/buttons.html}. We also recorded the overall orientation of the phone with respect to X, Y, Z directions. We used the device's geomagnetic field sensor in combination with the device's accelerometer to log those readings in \textit{Android Studio} logs.

\begin{enumerate}
    \item For X Direction, \textit{Pitch} is the angle between a plane parallel to the device's screen and a plane parallel to the ground. The range of values can be between -180 degrees to +180 degrees.
    \item For Y Direction, \textit{Roll} is the angle between a plane perpendicular to the device's screen and a plane. The range of values can be between -90 degrees to +90 degrees.
    \item For Z Direction, \textit{Azimuth} is the angle between the device's current compass direction and magnetic north. The range of values can be between 0 degrees to 270 degrees.
\end{enumerate}

\subsection{Apparatus}

Two different devices were used for the experiment - 

\begin{enumerate}
\item Google Pixel 3 (145.6 x 68.2 x 7.9 mm)
\item Motorola Moto G4 (153 x 76.6 x 9.8 mm)
\end{enumerate}

All information related to the devices have been collected from GSM Arena \footnote{https://www.gsmarena.com/}. Both devices run on the Android platform, with comparable specs and performance. Since our application (which has been designed for the purpose of this experiment) is not processor or memory heavy, we safely assumed that the minor individual differences in the smartphones won't influence any change in the data being gathered. No other active applications were running on the phone during the study. The phones were connected over WiFi to Android Studio and System logs were sent from the device to gather the data. To prevent other services of the phone, we used the device in airplane mode during our experiment. A mobile phone video camera was used to capture the complete study, focusing only on the device and user's hands.

\subsection{Variables}

\textbf{Independent Variables} were \emph{UI-Element Placement} (Baseline - top layout and adaptation - left/right layout), \emph{Interaction Hand} (dominant and non-dominant), \emph{Sub-Area} (1-6), and \emph{Phone Size} used (Google Pixel 3 and Motorola MotoG4). \textit{UI-Element Placement} denotes with which layout user interacted during the experiment. \textit{Interaction Hand} denotes whether the dominant or the non-dominant hand is being used during interaction. \textit{Sub-Area} denotes one of the 6 areas in which the Action Bar is divided in. \textit{Phone Size} used denotes with which device user interacts with during the experiment. 

\textbf{Dependent Variables} were \emph{Target Selection Time} [ms], \emph{Device Orientation} in X, Y, Z direction [degrees]. For each trial, \textit{Target Selection Time} denotes the time from the appearance of the instruction to choose a number on the screen till the time the target was actually hit. \textit{Device orientation} denotes the change in grip of the phone in the three dimensions. 

We recorded 3 UI-Element Placement (top/left/right) X 6 Sub-Area (1-6) X 2 Phone Size (Pixel/MotoG4) X 2 Interaction Hand (dominant/non-dominant) X 3 repetitions = 216 trials per participant. We also logged all trials to investigate potential outliers later. \emph{UI-Element Placement} and \emph{Phone Size} were both counterbalanced using a Latin Square, and Sub-Area selection sequence in all 3 layouts were randomized. Once all three repetitions were done, the user continued with the next technique, until all techniques had been tested. A trial session was presented before the start of the experiment to the user to get familiar with the task.

After the experiment concluded the users were asked to fill up a short Questionnaire where they were asked to rank the techniques based on perceived speed, safe grip and comfort.

\section{Results}
A total of 5 outliers were identified by applying the Tukey Method for Extreme Outliers on Time (7-12 sec). Looking at the video recordings for these trials revealed that users held the phone with both hands and then switched to one hand or were trying to adjust their sleeve/watch, while time was being counted, but they asked us to delete these button press. Hence, these trials were not representative and we excluded them from the analysis.

Firstly, we checked the normal distribution of our data using Q-Q plot and based on the result, we performed repeated measure multivariate ANOVA to report following results.

\textbf{Time Consumption:}
For non-dominant hand, the speed was statistically significantly different (F(2, 694) = 22.563, p < 0.05) for different placements of action bar. When comparing pairwise with Bonferroni post-hoc test, we found that for non-dominant hand left panel was significantly faster than both top panel (p = 0.0005) and right panel (p = 0.0005). \textit{Left (M = 1079.13 ms; 95\% CI [588.76, 1569.50]) incurred less time consumption than top (M = 1142.11 ms; 95\% CI [593.20, 1691.02]) and (M = 1378.05 ms; 95\% CI [578.926, 2177.174]) right panel. Difference between the means for left \& top is 62.98 ms and between right \& top is 298.92 ms.} For dominant hand, there was no significant result observed for any panel layout. However, the average time consumed for hitting target on left panel (1077.95 ms) was less than both top panel (1140.70 ms) and right panel (1369.76 ms). In addition, there was no significant effect of phone sizes on time required to hit the target.

\textbf{Device Orientation:}
The overall \textit{\textbf{pitch orientation angle}} was significantly different (F(2, 694) = 143.88, p < 0.05) for different layouts of the action bar in context of dominant hand. When comparing pairwise with Bonferroni post hoc test, we found that for dominant hand, right panel layout had  significantly less device pitch compared to left panel and top panel. \textit{Right (M = 0.4100 degrees ; 95\% CI [588.76, 1569.50]) incurred less pitch orientation angle than top (M = -1.7868 degrees; 95\% CI [1.830, 2.564]) and (M = -1.5609 degrees; 95\% CI [1.628, 2.314]) left panel. Difference between the means for right \& top is 2.197 degrees and between right \& left is 1.971 degrees.} However, this significant difference was not found for non-significant hand.

For device \textit{\textbf{roll orientation angle}}, we found both left and right layout having less device roll compared to top panel (F(2, 1378) = 406.976, p < 0.05). \textit{Right (M = 2.773 degrees ; 95\% CI [2.597, 2.951]) and left (M = 2.619 degrees; 95\% CI [2.411, 2.828]) incurred less pitch orientation angle than top (M = 4.376 degrees; 95\% CI [4.240, 4.514]). Difference between the means for top \& right is 1.603 degrees and between top \& left is 1.757 degrees.} However, there was no significant difference between left and right layout. In addition, it was interesting to note that the right panel layout also had significantly less device roll (p=0.0001) when compared to left panel layout. In all, these results had no significant impact for dominant and non-dominant hands \& both phone sizes.

We found an interesting result for \textit{\textbf{azimuth device orientation angle}} for dominant hand. When performing paired t-test on our combinations we found that the top panel layout significantly incurred less azimuth orientation t(347) =  23.570, p < 0.0005 compared to right layout. \textit{Top (M = 7.713 degrees; 95\% CI [6.289, 9.137]) incurred less azimuth orientation than (M = 8.819 degrees; 95\% CI [7.708, 9.93]) right panel. Difference between the means is 1.106 degrees.} This implies more safer grip of the phone with action bar being on left panel of the device. The result was consistent across both the phone sizes.

\textbf{Far Targets:}
We also found positive correlations between time taken and the distance to the far targets. We define far targets as targets on the screen edges i.e. 1 and 6 which are not ease to reach. On comparing left and right screen layout of the action bar for non-dominant hand, we found right screen layout has significant positive co-relation (r=0.342, p<0.001) with time taken to hit far targets compared to left screen layout. However, there was no significant difference between top layout compared to both right and left layout. Similarly, there was no significant difference when we take into account the dominant hand.

%\textbf{Success:} 5 out of 8 participants achieved 100\% success while hitting the targets on the action bar for the complete experiment. 2 participants scored 96.2\% \& 95.1\% accuracy. For these participants all the errors were recorded for their non-dominant hand trials while hitting the far targets only. 1 participant had a accuracy of 91.3\%. This participant had 4 errors while interacting with non-dominant hand for far targets and 2 errors while interacting with dominant hand for left version of adapted UI. There were neither significant main effects nor interaction effects for Success.

\begin{figure}[H]
\centering
  \includegraphics[width=1.1\columnwidth]{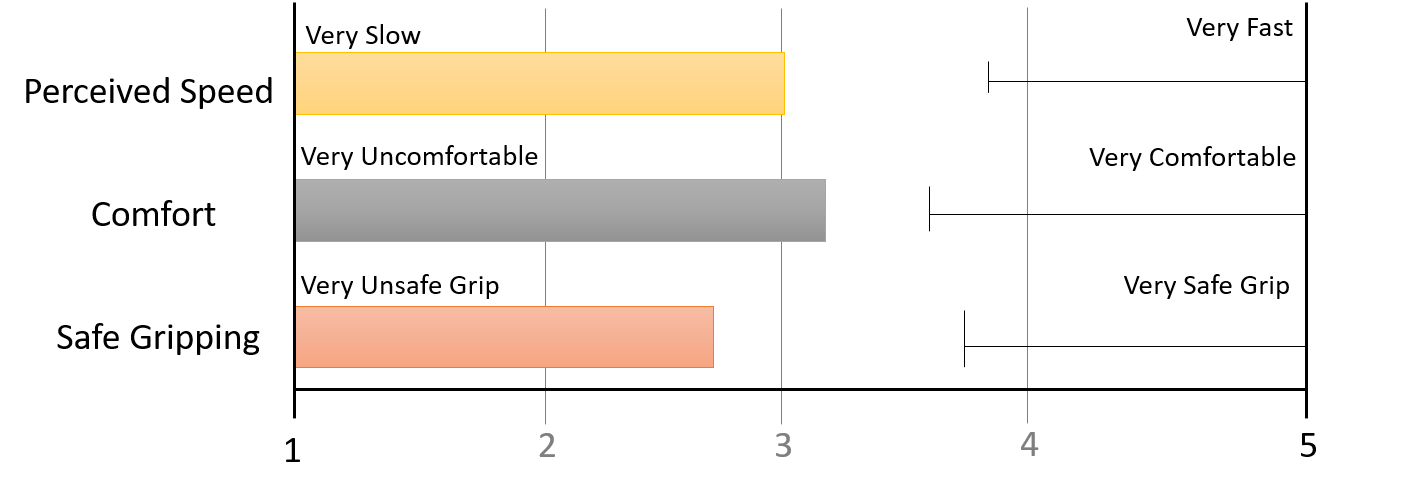}
   \caption{The chart show user feedback (dominant hand) (P=8) where 5 = very fast,
very fast and very comfortable, suggesting that users found the \textbf{baseline panel}
less comfortable and had unsafe gripping as per their experience. Perceived Speed (mean = 3.055, std = 1.198), Comfort (mean
= 3.364, std = 1.249), and Safe Gripping (mean = 2.733, std = 1.290).}~\label{fig:q-top}
\end{figure}
\begin{figure}[H]
\centering
  \includegraphics[width=1.1\columnwidth]{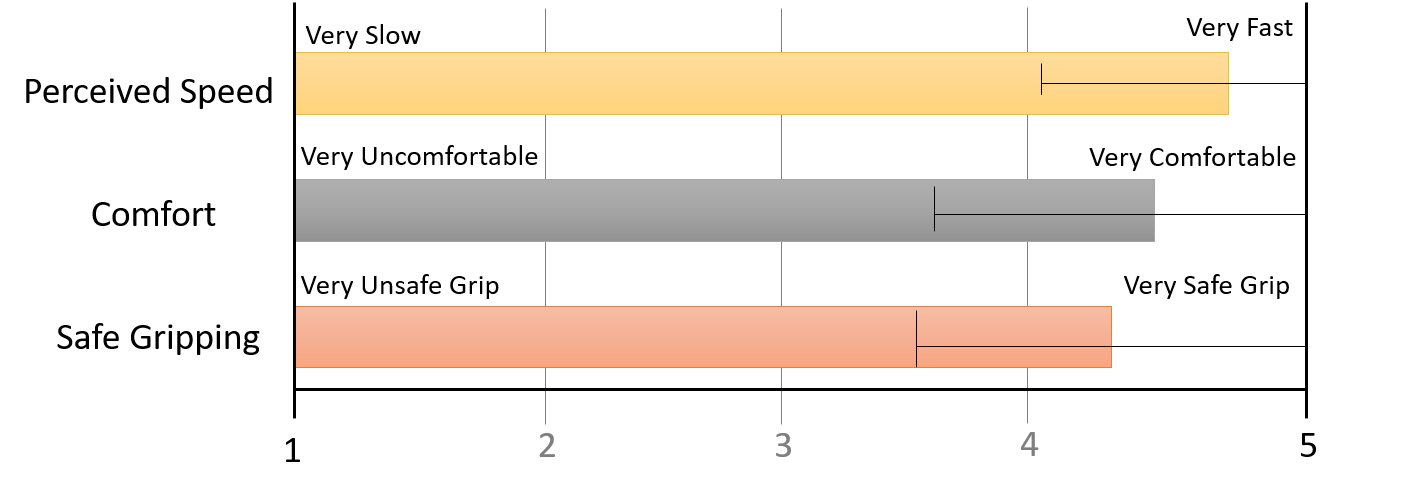}
   \caption{The chart show user feedback (dominant hand) (P=8) where 5 = very fast,
very fast and very comfortable, suggesting that users found the \textbf{adaptation panel}
easy to use and were satisfied overall with the experience. Perceived Speed (mean = 4.564, std = 0.798), Comfort (mean
= 4.464, std = 1.265), and Safe Gripping (mean = 4.235, std = 1.361).}~\label{fig:q-left-right}
\end{figure}

\textbf{User Ratings:}
We recorded Likert scale data for user ratings. Adaptation elements were perceived as faster, more comfortable to use, and more grip safe compared to the baseline top position for use with dominant hand for both the devices (refer figure \ref{fig:q-top}). On the other hand, the baseline layout 'top' position was considerably less fast, comfortable and had unstable grip compared to adaptation condition (refer figure  \ref{fig:q-left-right}). Given the small sample size, we refrain from reporting any statistical analyses.

\textbf{Video Recording Analysis}
We attempted to find out if there exists any patterns while holding device and hitting targets on the action bar. We found 4 different patterns that were common across all of the participants.
\begin{figure}[H]
\centering
  \includegraphics[width=1.0\columnwidth]{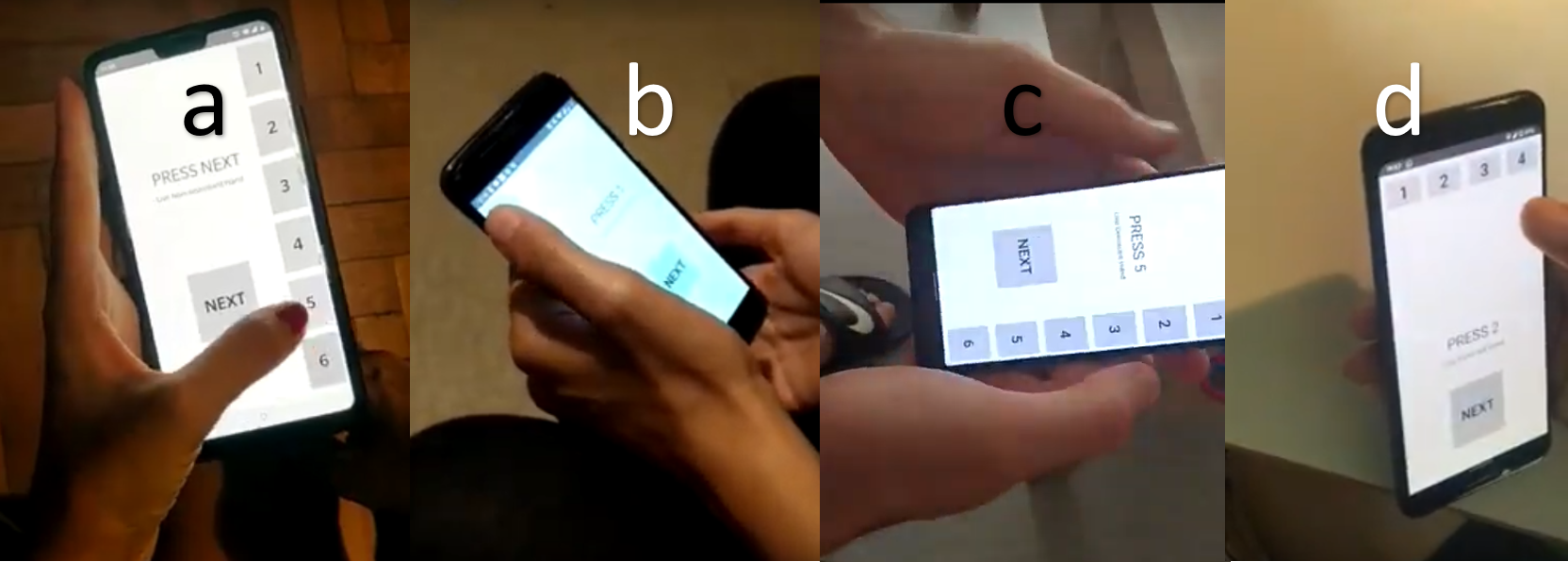}
   \caption{This figure shows 4 different observed patterns (a, b, c and d) while holding mobile devices when selecting targets on action bar.  a represents ease of use, b represents use of additional hand, c represents touching back side of phone and d represents change in grip}~\label{fig:grip-patterns}
\end{figure}
In figure \ref{fig:grip-patterns}a, for a left-handed user the adapted layout on left and right panel of the screen was easy to navigate in terms of change of gripping of the phone compared to baseline layout of top panel. For figure \ref{fig:grip-patterns}b, some of our right-handed users placed their right hand under the phone while performing the non-dominant hand trial in order to make sure phone doesn't fall down. In figure \ref{fig:grip-patterns}c, a right-handed user placed left hand on back side of phone to change his grip. Another interesting pattern was found for baseline layout. In figure \ref{fig:grip-patterns}d, a right-handed user had to change her grip always whenever she tried reaching the buttons on the top version of the action bar.

\section{Discussion}
Our results show how users interact with touch screen mobile devices for action bar. Action bar is a vital UI element in almost all the applications used independent of the device platform. Our study can help designers to consider hand dominance and UI adaptations while designing the action bar. We first discuss the importance of one-handed use of mobile phones and then draw the reader's attention to already available techniques and our difference with them through experiment.

We then discuss our experiment focusing upon 2 different devices with 3 different layouts for action bar including the existing use of action bar i.e. at the top of the screen. Our results suggest that hand dominance plays a crucial role while approaching targets on the action bar. We however didn't find any significant difference in our adaptation compared to existing position in terms of time consumption. The difference in means for time in both condition waves a path that researchers can consider as potential research area in future. We found significant differences in pitch and azimuth angles of orientation for our designed adaptation on comparing with top layout. These results infers higher stability of phone while reaching far targets so that users have better grip. In addition we found a positive co-relation between far targets and time consumption in all the 3 techniques used in the experiment. 

Our user ratings indicates that designed adaptations were perceived as faster, more comfortable and offering more stability for the smartphone. These results are at par with the observed logged data, however the recorded data didn't reveal any significant result. In addition, our video recording analyses reveals interesting observed patterns. We found left-handed users doesn't change their grip as much in our UI adaptations compared to baseline top layout. Also, there is a common pattern to hold the mobile device to save it from falling. Either users place their free hand beneath the phone while interacting with action bar or they firmly hold the phone with strength so that it doesn't fall down.

As expected for right-handed users placing action bar on right side of the screen would be easy to reach. Naturally, the displacement between the target and thumb position is decreased owing to faster hitting of target. Nevertheless, the top layout was user's most experienced way of interacting with action bar, we found adaptations to be useful as well. In all, significant results for orientation led us to investigate whether we could now effectively consider other UI elements in our future designs.

\section{Limitations and Future Work}
In our study, we only tested our adaptations only for action bar. There are other important elements in devices which plays crucial role for reachability as well. Unlike testing mere speed we incorporated several factors which can influence the target hitting at far areas on the mobile touch screen. We also displayed only 6 sub-areas of action bar which were essential for the covering the entire bar area. This was done in order to utilize maximum screen area. However, the sub-area count can be less or more for different screen sizes. Our future work would involve investigating UI adaptations for floating action bar, tabbed menus and hamburger menu options. We also aim to conduct a comparative future study in a controlled environment v/s in different scenarios in field to better understand how users use adaptations when we can automatically switch layout based on users intentions.

\section{Summary and Conclusion}
In summary, we presented UI adaptations for action bar, an interaction technique to improve reachability for smartphones. With our study we found that UI adaptations were significantly better technique both in terms of speed and device orientation if users use non-dominant hand to reach targets on action bar. It offers better grip and stability of smartphones. No significant difference was found for use of UI adaptations compared to existing layout of action bar for time consumption i.e. at the top of the screen. Our users rated designed adaptations faster, comfortable to use and offering stable grip when compared with existing layout. In addition, video analyses shows that with designed adaptations stable grip coping strategies were better while using action bar one handed.

\section{Acknowledgement}
The authors would like to extend warm thanks to all participants who participated in this study. Also, the authors thank Marcel \& Oliver for helping them in designing the study and providing feedback on the overall project.
\bibliographystyle{ACM-Reference-Format}
\bibliography{sample-authordraft}

\end{document}